\documentclass[sigconf]{acmart}
\AtBeginDocument{%
  }

\setcopyright{none}
\acmISBN{}
\acmDOI{}
\copyrightyear{}
\acmConference[CHI 2026 STAR Workshop]{Workshop on Science and Technology for Augmenting Reading at CHI 2026}{April 16, 2026}{Barcelona, Spain}

\makeatletter
\renewcommand{\@mkbibcitation}{\bgroup
  \let\@vspace\@vspace@orig
  \let\@vspacer\@vspacer@orig
  \def\@pages@word{\ifnum\getrefnumber{TotPages}=1\relax page\else pages\fi}%
  \def\footnotemark{}%
  \def\\{\unskip{} \ignorespaces}%
  \def\footnote{\ClassError{\@classname}{Please do not use footnotes
      inside a \string\title{} or \string\author{} command! Use
      \string\titlenote{} or \string\authornote{} instead!}}%
  \def\@article@string{\ifx\@acmArticle\@empty{\ }\else,
    Article~\@acmArticle\ \fi}%
  \par\medskip\small\noindent{\bfseries ACM Reference Format:}\par\nobreak
  \noindent\bgroup
    \def\\{\unskip{}, \ignorespaces}\authors\egroup. \@acmYear. \@title
  \ifx\@subtitle\@empty. \else: \@subtitle. \fi
  \if@ACM@nonacm\else
    \if@ACM@journal@bibstrip
       \textit{\@journalNameShort}
       \@acmVolume, \@acmNumber \@article@string (\@acmPubDate),
       \ref{TotPages}~\@pages@word.
    \else
       In \textit{\@acmBooktitle}%
       \ifx\@acmEditors\@empty\textit{.}\else
         \andify\@acmEditors\textit{, }\@acmEditors~\@editorsAbbrev.%
       \fi
    \fi
  \fi
  \ifx\@acmDOI\@empty\else\@formatdoi{\@acmDOI}\fi
\par\egroup}
\makeatother

\usepackage{todonotes}
\begin{document}

\title[Towards a Reader-Centred Taxonomy for Comprehension of AI Output]{From Binary Groundedness to Support Relations: Towards a Reader-Centred Taxonomy for Comprehension of AI Output}

\author{Advait Sarkar}
\affiliation{%
  \institution{Microsoft Research}
  \city{Cambridge}
  \country{United Kingdom}
}
\affiliation{%
  \institution{University of Cambridge}
  \city{Cambridge}
  \country{United Kingdom}
}
\affiliation{%
  \institution{University College London}
  \city{London}
  \country{United Kingdom}
}
\email{advait@microsoft.com}

\author{Christian Poelitz}
\affiliation{%
  \institution{Microsoft Research}
  \city{Cambridge}
  \country{United Kingdom}
}
\email{cpoelitz@microsoft.com}

\author{Viktor Kewenig}
\affiliation{%
  \institution{Microsoft Research}
  \city{Cambridge}
  \country{United Kingdom}
}
\email{a-vikewenig@microsoft.com}

\begin{abstract}
  Generative AI tools often answer questions using source documents, e.g., through retrieval augmented generation. Current groundedness and hallucination evaluations largely frame the relationship between an answer and its sources as binary (the answer is either supported or unsupported). However, this obscures both the syntactic moves (e.g., direct quotation vs. paraphrase) and the interpretive moves (e.g., induction vs. deduction) performed when models reformulate evidence into an answer. This limits both benchmarking and user-facing provenance interfaces. 
  
  We propose the development of a reader-centred taxonomy of grounding as a set of support relations between generated statements and source documents. We explain how this might be synthesised from prior research in linguistics and philosophy of language, and evaluated through a benchmark and human annotation protocol. Such a framework would enable interfaces that communicate not just whether a claim is grounded, but how.
\end{abstract}


\begin{teaserfigure}
  \centering
  \includegraphics[width=\textwidth]{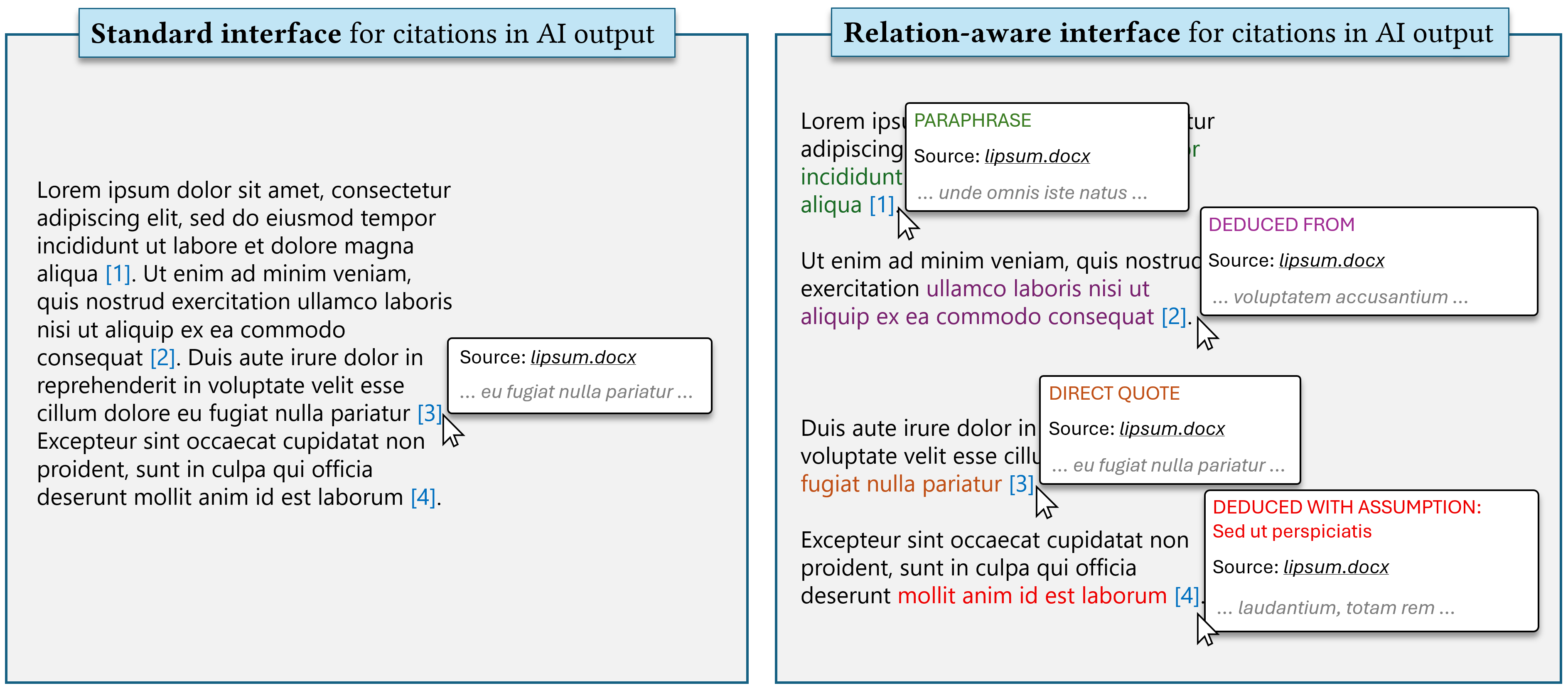}
  \caption{Left: standard citation-enabled responses from a language model. Right: a hypothetical interface that distinguishes between different types of syntactic manipulation (e.g. direct quote vs. paraphrase) and interpretation (e.g. induction, deduction, deduction subject to assumptions) involved in the production of language model output. We propose the development of a taxonomy of reader-centric \emph{support relations} that would enable such interfaces, thereby leading to better critical engagement of readers with language model output and understanding of how it relates to the sources.}
  \Description{The image is a side‑by‑side comparison of two user interface mock‑ups for how citations appear in AI‑generated text. The overall layout is a wide rectangle split vertically into two panels, with the left panel labeled “Standard interface for citations in AI output” and the right panel labeled “Relation‑aware interface for citations in AI output.” Both panels show blocks of placeholder text (“Lorem ipsum”) styled to look like paragraphs produced by a language model, with citation markers embedded in the text.
  In the left panel, the text appears as a continuous, plain paragraph in black, with small numbered citation markers in square brackets, such as “[1]”, “[2]”, “[3]”, and “[4]”, placed at the ends of sentences. A mouse cursor hovers over one of these citation numbers, triggering a small floating tooltip. That tooltip simply says “Source: lipsum.docx” followed by a short excerpt of text, indicating the cited source, but it does not explain how the AI used that source. All citations look visually identical and are treated the same way, regardless of whether the sentence is quoted, paraphrased, or inferred.
  In the right panel, the same general text content appears, but it is visually annotated to show different kinds of relationships between the AI’s output and its sources. Portions of the text are color‑coded, with different colors corresponding to different types of reasoning or transformation. When the mouse cursor hovers over a highlighted citation, a larger tooltip appears that includes a bold label describing the relationship. One tooltip is labeled “PARAPHRASE” in green, another is labeled “DEDUCED FROM” in purple, another is labeled “DIRECT QUOTE” in orange, and another is labeled “DEDUCED WITH ASSUMPTION” in red, with an example assumption shown in the tooltip text. Each tooltip also lists the source file name, again “lipsum.docx,” and shows a short excerpt from the source. The visual emphasis makes clear that different sentences or phrases are derived from sources in different ways, rather than all being treated as the same kind of citation.
  In summary: The image contrasts a flat, citation‑number‑only interface on the left with a richer, explanatory interface on the right that explicitly surfaces how the AI transformed or reasoned from its sources. The key visual message is that the right‑hand design gives readers more insight into whether text is quoted, paraphrased, or inferred, whereas the left‑hand design hides those distinctions behind uniform citation markers.}
  \label{fig:teaser}
\end{teaserfigure}

\maketitle

\section{Beyond Binary Hallucination and Groundedness}

When an Generative AI tool responds to a query by using information from a document or from the Web, ``groundedness'' is often treated as a binary property: either the answer is supported by a source, or it is not. Yet even the most straightforward fact retrieval scenario shows that this framing is too blunt.

Consider a short quarterly report paragraph about a fictional company (Acme Corporation), including the sentence ``Total revenue reached \$847.2 million…'' alongside surrounding narrative about performance and growth. Now imagine the user asks a concrete question tied to that excerpt: ``What was Acme Corp’s revenue in Q2 2025?''. When we tested this query on four commercial systems, they produced responses that were all plausibly ``grounded,'' but that differed materially in how they relate to the text: one repeated the exact figure verbatim (``\$847.2 million''); another quoted the figure but adds immediate surrounding context ``Total revenue reached \$847.2 million''; another rounds the figure (``Total revenue was over \$847 million''); and another paraphrases (``Revenue in Q2 2025 was US\$ 847.2 million'') in a way that more directly fits the query as asked, but quietly introduces an interpretive assumption (treating ``revenue'' as equivalent to ``total revenue,'' which may not match the user's interpretation). All of these responses are ``supported'' in the simple sense, but they represent different support relations; different ways of transforming source material into an answer.

The issue becomes unavoidable for interpretive questions. Using the same report-style excerpt, a user might ask: ``Is Acme doing well?'' Here, an answer cannot simply point to a single span; it must decide what counts as evidence (e.g., revenue growth percentage, or qualitative phrases that might appear in the report such as ``strong performance,'' ``resilience''), compress multiple sentences into a judgment, and reframe descriptive language as an evaluative conclusion. These are interpretive moves that embed implicit criteria for ``doing well'' and that must necessarily go beyond what the document literally supports.

Consider another toy example. If the source text is ``The cat sat on the mat'' and the user asks ``Did an animal sit on the mat?'', a ``yes'' answer relies on a background assumption (a cat is an animal). If the user asks ``Is the cat able to sit?'' the answer is a deduction from the described event. If the user asks ``What did the cat sit on?'' the answer can be a minimal direct quote (``a mat'') or a more complete quote (``The cat sat on the mat'') or a partial paraphrase (``It sat on the mat''). These responses differ in syntactic and epistemic status: direct quotation, paraphrases, deduction, and deduction contingent on ancillary assumptions, despite all being ``supported'' claims. Put differently, two answers can both be ``grounded'' while warranting very different levels of scrutiny: one may be traceable to a verbatim span; another may be a defensible but assumption-laden inference.

Prevailing groundedness frameworks that focus on binary ``supported/unsupported'' checks fail to model the syntactic and interpretive reformulation that occurs when text is turned into an answer to a user query. This matters because, as readers increasingly rely on Generative AI tools to process documents and synthesise answers for them, it becomes crucial for them to understand not merely whether support is present, but what kind of support.

Readers rarely have the time, expertise, or motivation to audit every claim in an AI-generated answer; by distinguishing these support relations, we can help readers assess where scrutiny is most warranted. Prior work has argued that the greatest risk of Generative AI to knowledge work is not hallucination, but the erosion of critical thinking through passive reliance~\cite{sarkar2024genAIcritical}, with survey evidence showing that higher confidence in AI is associated with less critical thinking and a shift from information gathering to verification~\cite{lee2025aisurvey}. Binary groundedness labels may exacerbate this: if a claim is labelled ``supported,'' readers may not interrogate \emph{how} it is supported. More broadly, AI has been argued to shift knowledge work from material production to the critical integration of AI output~\cite{sarkar2023aiknowledgework}, and the metacognitive demands this places on users (monitoring what the AI did, evaluating its output, and deciding how much to rely on it) have been identified as a key usability challenge~\cite{tankelevitch2024GenAImetacognition}. A taxonomy of support relations can be understood as a resource for developing metacognitive tools that helps readers meet these demands. Our design intent follows the argument that AI systems should challenge and provoke critical engagement rather than merely accelerate workflows~\cite{sarkar2024aiprovocateur}.

\section{Towards a Taxonomy of Support Relations}

Given a generated statement \(S\) and a source document (or set of documents) \(D\), we wish to determine a support relation (or relations) \(R(S,D)\) drawn from an explicitly defined inventory (e.g., direct quotation, paraphrase, deductive support, inductive support, support contingent on ancillary assumptions). How might we develop such an operational taxonomy of support relations, and how might we evaluate it? 

Prior work could provide strong theoretical starting points. For instance, Toulmin's model of argumentation \cite{Toulmin2003-xj,kneupper1978teaching} decomposes argumentative statements into claim, grounds/data, and warrant, with optional backing, qualifier, and rebuttal. Toulmin's scheme is attractive here because it renders explicit the inferential link between evidence and an assertion. A related starting point is work on support in argumentation systems \cite{Cohen2014survey}, which surveys distinct interpretations of support (e.g., deductive support, evidential support, necessary support, backing).

Another starting point is pragmatics. Grice's account of ``conversational implicature'' \cite{grice1991studies} describes how interpreters attribute speaker meaning beyond what is literally said, under cooperative principles and contextual assumptions. This suggests categories for content that is not entailed by \(D\) but is nevertheless inferable in context.

Finally, scholarship on representing scholarly discourse relations provides relevant inspiration for knowledge work settings: the ``ScholOnto'' project \cite{BuckinghamShum2000,ManciniBuckinghamShum2006,UrenBuckinghamShum2006,BuckinghamShum2007}
, for example, aims to support scholarly interpretation by enabling researchers to represent claims and their relationships to the literature as an explicit semantic network. The ``scite'' metric \cite{nicholson2021scite} aims to contextualise scientific citations as supportive or unsupportive.
 
Yet, it is not sufficient for a taxonomy to be theoretically grounded. It must also be practical on two counts: it must be possible for human and machine annotators to robustly classify support relations, and these support relations must be useful and understandable by readers of LLM-generated text. This could be achieved by treating taxonomy construction as an iterative design process, with explicit attention to operationalisation.

The initial step might be to conduct a structured literature review across the above traditions to produce a longlist of candidate support relations and boundary cases, followed by collapsing this longlist into a minimal working taxonomy guided by two pragmatic criteria: discriminability (can trained annotators reliably distinguish the categories) and actionability (does the distinction matter for downstream uses such as provenance interfaces). This could involve iterative reviews of the taxonomy by expert annotators (e.g. those familiar with linguistic theories of support) as well as non-expert readers.%

To quantitatively validate the taxonomy, the next step is to produce a full annotation specification, definitions, canonical examples, counterexamples, and decision rules, explicitly targeted at usability by both human annotators and potentially LLMs-as-judge~\cite{zheng2023judging} with the goal of scaling annotation to larger datasets and enabling future online annotation. A practical path is to construct a benchmark of statement-source pairs by enriching existing groundedness and hallucination corpora (some examples are given in Section~\ref{sec:prior-benchmarks}) with support-relation labels. Human reliability and construct validation could first be established with a human annotation study on a stratified sample of statement-source pairs. This would quantify inter-annotator agreement and reveal systematic confusions between relation types, enabling iterative refinement, and ideally result in a ``gold standard'' ground truth dataset of classified statement-source pairs. It would then be straightforward to benchmark the performance of frontier models on this annotation task by measuring their ability to match human annotations.

\section{Related Work}
\subsection{Science and Technology for Augmented Reading}
This project could build on prior HCI and design work in technologically-assisted reading. The Semantic Reader project \cite{lo2024semantic}, which encompasses much prior work in this area, outlines how citations in scientific papers can be enriched to support readers. For instance, inline citations can be coloured to indicate whether the reader has already recently encountered or read them, and can display personalised cards that explain how the work relates to their interests. The ``CiteRead'' \cite{rachatasumrit2022citeread} system integrates commentary from subsequently-published work to support evaluation of citations. 

The ``InkSync'' system \cite{laban2024inksync}, while not aimed at helping readers verify LLM text generated from sources, adopts a ``warn-verify-audit'' approach where LLM-generated text that appears to contain new information is highlighted as such, and the user is prompted to manually verify it (e.g., through a web search). Similarly, the ``GenAudit'' system \cite{krishna2025genaudit} identifies errors and suggests correct edit suggestions. Systems like these are helpful, but are focused on helping the user avoid factual errors, rather than evaluate the particular relationship of \emph{supported} claims to the source text, as we propose.

A relevant prior is the ``Traceable Texts'' interface \cite{kambhamettu2025traceable}, which annotates phrases in AI-generated summaries with links to corresponding phrases in the source, helping with fact checking and indexing into sources for deeper reading. The {\sc datatales} interface \cite{sultanum2023datatales} employs a similar brushing/linking interaction for textual narratives generated from data charts. Going further, the ``attribution gradients'' interface \cite{kambhamettu2025attribution} decomposes LLM-generated statements into claims, which are in turn linked to evidence sources, each of which is classified according to a $2\times2$ framework (first-degree vs. second-degree, support vs. contradiction). Another close precedent is the {\sc facts\&evidence} system \cite{boonsanong2025factsandevidence}, which helps users verify the factuality of a passage by decomposing it into claims and verifying the support for each claim against web sources. A related line of work proposes ``co-audit'' as a general framework for helping humans double-check AI-generated content, including design principles such as grounding outputs with sources and not allowing the LLM to audit itself~\cite{gordon2024coaudit}. Similarly, interactive task decomposition interfaces that surface editable assumptions and execution plans have been shown to improve users' ability to steer and verify AI-generated analyses~\cite{kazemitabaar2024steering}.

In the programming domain, the ``Trailblazer'' interface \cite{yan2025answering} helps developers understand LLM-generated answers about a codebase by visualising the trace of the agent's exploration of the codebase. This is an interesting approach because, as we will observe in the next section regarding much knowledge work, answers often do not draw on simple, single sources, but are distributed across a code (or more generally, knowledge) base. A somewhat similar interactive hierarchical exploration for academic papers is exemplified in the ``Qlarify'' \cite{fok2024qlarify} and ``TreeReader'' \cite{zhang2025treereader} interfaces.

\subsection{Hallucination Detection and Groundedness Benchmarks}
\label{sec:prior-benchmarks}
There is a vast literature on hallucination detection and various benchmarks that we cannot exhaustively review here (a review is given by \citet{kazlaris2025hallusurvey}). Prominent examples include HaluEval \cite{li2023halueval}, HaDes \cite{liu2022hades}, FactCHD \cite{chen2025factchd}, RAGTruth \cite{niu2024ragtruth}, DiaHalu \cite{chen2024diahalu}, and Hallulens \cite{bang2025hallulens}. These are noted as potential starting points, since these datasets are often composed of source-statement pairs annotated with binary labels, that we could expand with a richer set of support relation labels.

Prior research has gone beyond binary categorisations, albeit not in the direction of classifying different types of positive support as we have proposed here. Rather, here the focus is in understanding different types of errors. For instance, the FRANK benchmark \cite{pagnoni2021understanding} identifies a typology of seven errors, such as relation error, entity error, out-of-article error, and grammatical errors. Similarly, the LibreEval dataset \cite{arize_libreeval} identifies six hallucination subtypes. The FActScore benchmark \cite{min2023factscore} uses a ternary top level categorisation: supported, not supported and irrelevant; however in a qualitative analysis they note finer-grained types pertaining to the scope of the error, such as single-sentence contradiction and page-level contradiction, as well as pertaining to the semantic manipulation, such as ``subjective'' judgements, and cases where the source is itself wrong or inconsistent. Similarly, AttrScore \cite{yue2023attrscore} uses a ternary classification: the statement is either attributable (i.e., supported), extrapolatory (i.e., unsupported), or contradictory.

The Claimify method \cite{metropolitansky2025towards} notes that identifying how LLM-generated statements map onto knowledge claims is itself a complex and nonstandard procedure, and the method proposed can extract atomic claims from complex sentences, and identify amiguous claims that cannot be resolved. The claims so extracted can then be used as part of a verification pipeline such as VeriTrail \cite{metropolitansky2025veritrail}. It may also be possible to frontload the identification of claim spans at generation time, using a technique such as symbolically grounded generation \cite{hennigen2024symgen} or citation-enabled LLMs (a review is given by \citet{gao2023alce}).

\section{Discussion and Open Questions}

There is a real tension between a taxonomy that is theoretically justified and one that is operationally usable at scale. Many candidate relation types depend on background knowledge and context, and the family of relations that involve support contingent on ancillary assumptions can proliferate rapidly if not constrained, because almost any inference can be described as requiring a tacit premise. Any attempt to import the full richness of scholarly discourse on argumentation, linguistics, and philosophy of language risks producing an ontology that is too fine-grained for annotators to apply consistently, or too abstract to be useful to readers as decision support. This implies an open design question: what is the smallest set of support relations that remains meaningful for annotators and readers, remaining theoretically advised by but without ``overfitting'' to philosophically motivated distinctions?

There are basic questions of the ``units'' of analysis: we have proposed defining the task as relating a generated statement \(S\) to evidence in a document \(D\), but in practice it is rarely obvious what should count as a single statement, how to segment complex sentences, or when to decompose an answer into smaller atomic propositions. Existing pipelines highlight both opportunity and hazard: atomic decomposition can make verification tractable, but it introduces a consequential modelling choice about how fine-grained the decomposition should be and what contextual information must remain attached so that the unit remains faithful and interpretable \cite{metropolitansky2025towards}. For a support-relations taxonomy, this raises an open question about the division of labour between claim extraction and support labelling.

Similarly, many answers are supported not by a single contiguous span, but by multiple excerpts distributed across a document, and sometimes by excerpts that appear to pull in different directions. Real knowledge work is rarely partitioned into isolated documents: what matters is often a cloud of concepts, commitments, themes, and discussions distributed across conversations and artefacts, with users wanting to trace where else something was discussed and how it connects to other work. This raises questions about how to scale the representation when the graph of relevant materials becomes larger. Our design intent is not for the reader's evaluation process to fall back on ``the AI knows and tells me,'' but to preserve users’ ability to know where statements come from and to dive into the right parts of the underlying material. For the benchmark and annotation scheme, this implies open questions about whether the support relation should be defined between \(S\) and a single excerpt, between \(S\) and a set of excerpts, or between \(S\) and a structured evidential object that can represent corroboration and tension within \(D\) and across $D_1,\dots,D_n$. It also forces an explicit choice about how to treat internal inconsistency: whether contradictory support should be surfaced to readers as a distinct relation, a meta-property of the evidence set, or discursively within the generated statement itself.

Another open issue is domain transfer. Support relations may manifest differently across summarisation, document-grounded question answering, and more analytical or evaluative queries, and even knowledge work domains (e.g., legal work, medical work, scientific research, etc.) because the permissible interpretive moves, the expected level of compression, and the availability of an agreed evidential standard vary by task and domain. We must also be sensitive to the fragilities of the LLM-as-judge paradigm: model behaviour and evaluation can depend strongly on task framing and prompts. Recent work has shown that even highly detailed rubric instructions yield only marginal improvements in LLM-as-judge alignment with human judgements, and that simpler measures such as perplexity can sometimes perform comparably~\cite{murugadoss2025evaluating}. Moreover, LLM-generated ``explanations'' of their own reasoning have been shown to be unreliable, since they do not reflect the model's actual mechanism~\cite{sarkar2024llmscannotexplain}.

We close by inviting collaboration precisely because we believe these to be productive research questions. The project calls for interdisciplinary input on which theoretical frameworks offer the best starting points for an operational taxonomy. We particularly welcome discussion of alternative choices about claim granularity and decomposition, and discussion on principled ways to represent support when evidence is distributed across multiple spans or multiple documents, including cases where evidence is cumulative or internally in tension. Finally, we hope to discuss benchmarking best practices and prompt design, including protocols that make prompt sensitivity and task transfer explicit. The objective is to build support for understanding provenance and critical reading in a form that remains legible and actionable for end users.

\bibliographystyle{ACM-Reference-Format}
\bibliography{references}


\begin{thebibliography}{45}


\ifx \showCODEN    \undefined \def \showCODEN     #1{\unskip}     \fi
\ifx \showISBNx    \undefined \def \showISBNx     #1{\unskip}     \fi
\ifx \showISBNxiii \undefined \def \showISBNxiii  #1{\unskip}     \fi
\ifx \showISSN     \undefined \def \showISSN      #1{\unskip}     \fi
\ifx \showLCCN     \undefined \def \showLCCN      #1{\unskip}     \fi
\ifx \shownote     \undefined \def \shownote      #1{#1}          \fi
\ifx \showarticletitle \undefined \def \showarticletitle #1{#1}   \fi
\ifx \showURL      \undefined \def \showURL       {\relax}        \fi
\providecommand\bibfield[2]{#2}
\providecommand\bibinfo[2]{#2}
\providecommand\natexlab[1]{#1}
\providecommand\showeprint[2][]{arXiv:#2}

\bibitem[{Arize AI}({[n.\,d.]})]%
        {arize_libreeval}
\bibfield{author}{\bibinfo{person}{{Arize AI}}.} \bibinfo{year}{[n.\,d.]}\natexlab{}.
\newblock \bibinfo{title}{LibreEval: The Open-Source Benchmark for {RAG} Hallucination Detection}.
\newblock \bibinfo{howpublished}{\url{https://arize.com/llm-hallucination-dataset/}}.
\newblock
\newblock
\shownote{Accessed 4 February 2026}.


\bibitem[Bang et~al\mbox{.}(2025)]%
        {bang2025hallulens}
\bibfield{author}{\bibinfo{person}{Yejin Bang}, \bibinfo{person}{Ziwei Ji}, \bibinfo{person}{Alan Schelten}, \bibinfo{person}{Anthony Hartshorn}, \bibinfo{person}{Tara Fowler}, \bibinfo{person}{Cheng Zhang}, \bibinfo{person}{Nicola Cancedda}, {and} \bibinfo{person}{Pascale Fung}.} \bibinfo{year}{2025}\natexlab{}.
\newblock \showarticletitle{Hallulens: Llm hallucination benchmark}.
\newblock \bibinfo{journal}{\emph{arXiv preprint arXiv:2504.17550}} (\bibinfo{year}{2025}).
\newblock


\bibitem[Boonsanong et~al\mbox{.}(2025)]%
        {boonsanong2025factsandevidence}
\bibfield{author}{\bibinfo{person}{Varich Boonsanong}, \bibinfo{person}{Vidhisha Balachandran}, \bibinfo{person}{Xiaochuang Han}, \bibinfo{person}{Shangbin Feng}, \bibinfo{person}{Lucy~Lu Wang}, {and} \bibinfo{person}{Yulia Tsvetkov}.} \bibinfo{year}{2025}\natexlab{}.
\newblock \showarticletitle{{FACTS}{\&}{EVIDENCE}: An Interactive Tool for Transparent Fine-Grained Factual Verification of Machine-Generated Text}. In \bibinfo{booktitle}{\emph{Proceedings of the 2025 Conference of the Nations of the Americas Chapter of the Association for Computational Linguistics: Human Language Technologies (System Demonstrations)}}, \bibfield{editor}{\bibinfo{person}{Nouha Dziri}, \bibinfo{person}{Sean~(Xiang) Ren}, {and} \bibinfo{person}{Shizhe Diao}} (Eds.). \bibinfo{publisher}{Association for Computational Linguistics}, \bibinfo{address}{Albuquerque, New Mexico}, \bibinfo{pages}{437--448}.
\newblock
\showISBNx{979-8-89176-191-9}
\href{https://doi.org/10.18653/v1/2025.naacl-demo.35}{doi:\nolinkurl{10.18653/v1/2025.naacl-demo.35}}


\bibitem[Buckingham~Shum et~al\mbox{.}(2000)]%
        {BuckinghamShum2000}
\bibfield{author}{\bibinfo{person}{Simon Buckingham~Shum}, \bibinfo{person}{Enrico Motta}, {and} \bibinfo{person}{John Domingue}.} \bibinfo{year}{2000}\natexlab{}.
\newblock \showarticletitle{ScholOnto: an ontology-based digital library server for research documents and discourse}.
\newblock \bibinfo{journal}{\emph{International Journal on Digital Libraries}} \bibinfo{volume}{3}, \bibinfo{number}{3} (\bibinfo{year}{2000}), \bibinfo{pages}{237--248}.
\newblock
\href{https://doi.org/10.1007/s007990000034}{doi:\nolinkurl{10.1007/s007990000034}}


\bibitem[Buckingham~Shum et~al\mbox{.}(2007)]%
        {BuckinghamShum2007}
\bibfield{author}{\bibinfo{person}{Simon~J. Buckingham~Shum}, \bibinfo{person}{Victoria Uren}, \bibinfo{person}{Gangmin Li}, \bibinfo{person}{Bertrand Sereno}, {and} \bibinfo{person}{Clara Mancini}.} \bibinfo{year}{2007}\natexlab{}.
\newblock \showarticletitle{Modelling naturalistic argumentation in research literatures: representation and interaction design issues}.
\newblock \bibinfo{journal}{\emph{International Journal of Intelligent Systems}} \bibinfo{volume}{22}, \bibinfo{number}{1} (\bibinfo{year}{2007}), \bibinfo{pages}{17--47}.
\newblock
\href{https://doi.org/10.1002/int.20188}{doi:\nolinkurl{10.1002/int.20188}}


\bibitem[Chen et~al\mbox{.}(2024a)]%
        {chen2024diahalu}
\bibfield{author}{\bibinfo{person}{Kedi Chen}, \bibinfo{person}{Qin Chen}, \bibinfo{person}{Jie Zhou}, \bibinfo{person}{He Yishen}, {and} \bibinfo{person}{Liang He}.} \bibinfo{year}{2024}\natexlab{a}.
\newblock \showarticletitle{Diahalu: A dialogue-level hallucination evaluation benchmark for large language models}. In \bibinfo{booktitle}{\emph{Findings of the Association for Computational Linguistics: EMNLP 2024}}. \bibinfo{pages}{9057--9079}.
\newblock


\bibitem[Chen et~al\mbox{.}(2024b)]%
        {chen2025factchd}
\bibfield{author}{\bibinfo{person}{Xiang Chen}, \bibinfo{person}{Duanzheng Song}, \bibinfo{person}{Honghao Gui}, \bibinfo{person}{Chenxi Wang}, \bibinfo{person}{Ningyu Zhang}, \bibinfo{person}{Yong Jiang}, \bibinfo{person}{Fei Huang}, \bibinfo{person}{Chengfei Lyu}, \bibinfo{person}{Dan Zhang}, {and} \bibinfo{person}{Huajun Chen}.} \bibinfo{year}{2024}\natexlab{b}.
\newblock \showarticletitle{FactCHD: benchmarking fact-conflicting hallucination detection}. In \bibinfo{booktitle}{\emph{Proceedings of the Thirty-Third International Joint Conference on Artificial Intelligence}} (Jeju, Korea) \emph{(\bibinfo{series}{IJCAI '24})}. Article \bibinfo{articleno}{687}, \bibinfo{numpages}{9}~pages.
\newblock
\showISBNx{978-1-956792-04-1}
\href{https://doi.org/10.24963/ijcai.2024/687}{doi:\nolinkurl{10.24963/ijcai.2024/687}}


\bibitem[Cohen et~al\mbox{.}(2014)]%
        {Cohen2014survey}
\bibfield{author}{\bibinfo{person}{Andrea Cohen}, \bibinfo{person}{Sebastian Gottifredi}, \bibinfo{person}{Alejandro~J. García}, {and} \bibinfo{person}{Guillermo~R. Simari}.} \bibinfo{year}{2014}\natexlab{}.
\newblock \showarticletitle{A survey of different approaches to support in argumentation systems}.
\newblock \bibinfo{journal}{\emph{The Knowledge Engineering Review}} \bibinfo{volume}{29}, \bibinfo{number}{5} (\bibinfo{year}{2014}), \bibinfo{pages}{513–550}.
\newblock
\href{https://doi.org/10.1017/S0269888913000325}{doi:\nolinkurl{10.1017/S0269888913000325}}


\bibitem[Fok et~al\mbox{.}(2024)]%
        {fok2024qlarify}
\bibfield{author}{\bibinfo{person}{Raymond Fok}, \bibinfo{person}{Joseph~Chee Chang}, \bibinfo{person}{Tal August}, \bibinfo{person}{Amy~X. Zhang}, {and} \bibinfo{person}{Daniel~S. Weld}.} \bibinfo{year}{2024}\natexlab{}.
\newblock \showarticletitle{Qlarify: Recursively Expandable Abstracts for Dynamic Information Retrieval over Scientific Papers}. In \bibinfo{booktitle}{\emph{Proceedings of the 37th Annual ACM Symposium on User Interface Software and Technology}} (Pittsburgh, PA, USA) \emph{(\bibinfo{series}{UIST '24})}. \bibinfo{publisher}{Association for Computing Machinery}, \bibinfo{address}{New York, NY, USA}, Article \bibinfo{articleno}{145}, \bibinfo{numpages}{21}~pages.
\newblock
\showISBNx{9798400706288}
\href{https://doi.org/10.1145/3654777.3676397}{doi:\nolinkurl{10.1145/3654777.3676397}}


\bibitem[Gao et~al\mbox{.}(2023)]%
        {gao2023alce}
\bibfield{author}{\bibinfo{person}{Tianyu Gao}, \bibinfo{person}{Howard Yen}, \bibinfo{person}{Jiatong Yu}, {and} \bibinfo{person}{Danqi Chen}.} \bibinfo{year}{2023}\natexlab{}.
\newblock \showarticletitle{Enabling Large Language Models to Generate Text with Citations}. In \bibinfo{booktitle}{\emph{Proceedings of the 2023 Conference on Empirical Methods in Natural Language Processing}}, \bibfield{editor}{\bibinfo{person}{Houda Bouamor}, \bibinfo{person}{Juan Pino}, {and} \bibinfo{person}{Kalika Bali}} (Eds.). \bibinfo{publisher}{Association for Computational Linguistics}, \bibinfo{address}{Singapore}, \bibinfo{pages}{6465--6488}.
\newblock
\href{https://doi.org/10.18653/v1/2023.emnlp-main.398}{doi:\nolinkurl{10.18653/v1/2023.emnlp-main.398}}


\bibitem[Gordon et~al\mbox{.}(2024)]%
        {gordon2024coaudit}
\bibfield{author}{\bibinfo{person}{Andrew~D. Gordon}, \bibinfo{person}{Carina Negreanu}, \bibinfo{person}{José Cambronero}, \bibinfo{person}{Rasika Chakravarthy}, \bibinfo{person}{Ian Drosos}, \bibinfo{person}{Hao Fang}, \bibinfo{person}{Bhaskar Mitra}, \bibinfo{person}{Hannah Richardson}, \bibinfo{person}{Advait Sarkar}, \bibinfo{person}{Stephanie Simmons}, \bibinfo{person}{Jack Williams}, {and} \bibinfo{person}{Ben Zorn}.} \bibinfo{year}{2024}\natexlab{}.
\newblock \showarticletitle{{Co-audit: tools to help humans double-check AI-generated content}}.
\newblock \bibinfo{journal}{\emph{Proceedings of the 14th annual workshop on the intersection of HCI and PL (PLATEAU 2024)}} (\bibinfo{date}{5} \bibinfo{year}{2024}).
\newblock
\href{https://doi.org/10.1184/R1/25587552.v1}{doi:\nolinkurl{10.1184/R1/25587552.v1}}


\bibitem[Grice(1991)]%
        {grice1991studies}
\bibfield{author}{\bibinfo{person}{Paul Grice}.} \bibinfo{year}{1991}\natexlab{}.
\newblock \bibinfo{booktitle}{\emph{Studies in the Way of Words}}.
\newblock \bibinfo{publisher}{Harvard University Press}.
\newblock


\bibitem[Hennigen et~al\mbox{.}(2024)]%
        {hennigen2024symgen}
\bibfield{author}{\bibinfo{person}{Lucas~Torroba Hennigen}, \bibinfo{person}{Shannon Shen}, \bibinfo{person}{Aniruddha Nrusimha}, \bibinfo{person}{Bernhard Gapp}, \bibinfo{person}{David Sontag}, {and} \bibinfo{person}{Yoon Kim}.} \bibinfo{year}{2024}\natexlab{}.
\newblock \bibinfo{title}{Towards Verifiable Text Generation with Symbolic References}.
\newblock
\showeprint[arxiv]{2311.09188}~[cs.CL]
\urldef\tempurl%
\url{https://arxiv.org/abs/2311.09188}
\showURL{%
\tempurl}


\bibitem[Kambhamettu et~al\mbox{.}(2025a)]%
        {kambhamettu2025traceable}
\bibfield{author}{\bibinfo{person}{Hita Kambhamettu}, \bibinfo{person}{Jamie Flores}, {and} \bibinfo{person}{Andrew Head}.} \bibinfo{year}{2025}\natexlab{a}.
\newblock \showarticletitle{Traceable Texts and Their Effects: A Study of Summary-Source Links in AI-Generated Summaries}. In \bibinfo{booktitle}{\emph{Proceedings of the Extended Abstracts of the CHI Conference on Human Factors in Computing Systems}} \emph{(\bibinfo{series}{CHI EA '25})}. \bibinfo{publisher}{Association for Computing Machinery}, \bibinfo{address}{New York, NY, USA}, Article \bibinfo{articleno}{538}, \bibinfo{numpages}{7}~pages.
\newblock
\showISBNx{9798400713958}
\href{https://doi.org/10.1145/3706599.3719830}{doi:\nolinkurl{10.1145/3706599.3719830}}


\bibitem[Kambhamettu et~al\mbox{.}(2025b)]%
        {kambhamettu2025attribution}
\bibfield{author}{\bibinfo{person}{Hita Kambhamettu}, \bibinfo{person}{Alyssa Hwang}, \bibinfo{person}{Philippe Laban}, {and} \bibinfo{person}{Andrew Head}.} \bibinfo{year}{2025}\natexlab{b}.
\newblock \bibinfo{title}{Attribution Gradients: Incrementally Unfolding Citations for Critical Examination of Attributed AI Answers}.
\newblock
\showeprint[arxiv]{2510.00361}~[cs.HC]
\urldef\tempurl%
\url{https://arxiv.org/abs/2510.00361}
\showURL{%
\tempurl}


\bibitem[Kazemitabaar et~al\mbox{.}(2024)]%
        {kazemitabaar2024steering}
\bibfield{author}{\bibinfo{person}{Majeed Kazemitabaar}, \bibinfo{person}{Jack Williams}, \bibinfo{person}{Ian Drosos}, \bibinfo{person}{Tovi Grossman}, \bibinfo{person}{Austin~Zachary Henley}, \bibinfo{person}{Carina Negreanu}, {and} \bibinfo{person}{Advait Sarkar}.} \bibinfo{year}{2024}\natexlab{}.
\newblock \showarticletitle{Improving Steering and Verification in AI-Assisted Data Analysis with Interactive Task Decomposition}. In \bibinfo{booktitle}{\emph{Proceedings of the 37th Annual ACM Symposium on User Interface Software and Technology}} (Pittsburgh, PA, USA) \emph{(\bibinfo{series}{UIST '24})}. \bibinfo{publisher}{Association for Computing Machinery}, \bibinfo{address}{New York, NY, USA}, Article \bibinfo{articleno}{92}, \bibinfo{numpages}{19}~pages.
\newblock
\showISBNx{9798400706288}
\href{https://doi.org/10.1145/3654777.3676345}{doi:\nolinkurl{10.1145/3654777.3676345}}


\bibitem[Kazlaris et~al\mbox{.}(2025)]%
        {kazlaris2025hallusurvey}
\bibfield{author}{\bibinfo{person}{Ioannis Kazlaris}, \bibinfo{person}{Efstathios Antoniou}, \bibinfo{person}{Konstantinos Diamantaras}, {and} \bibinfo{person}{Charalampos Bratsas}.} \bibinfo{year}{2025}\natexlab{}.
\newblock \showarticletitle{From Illusion to Insight: A Taxonomic Survey of Hallucination Mitigation Techniques in LLMs}.
\newblock \bibinfo{journal}{\emph{AI}} \bibinfo{volume}{6}, \bibinfo{number}{10} (\bibinfo{year}{2025}).
\newblock
\showISSN{2673-2688}
\href{https://doi.org/10.3390/ai6100260}{doi:\nolinkurl{10.3390/ai6100260}}


\bibitem[Kneupper(1978)]%
        {kneupper1978teaching}
\bibfield{author}{\bibinfo{person}{Charles~W Kneupper}.} \bibinfo{year}{1978}\natexlab{}.
\newblock \showarticletitle{Teaching argument: An introduction to the Toulmin model}.
\newblock \bibinfo{journal}{\emph{College Composition \& Communication}} \bibinfo{volume}{29}, \bibinfo{number}{3} (\bibinfo{year}{1978}), \bibinfo{pages}{237--241}.
\newblock


\bibitem[Krishna et~al\mbox{.}(2025)]%
        {krishna2025genaudit}
\bibfield{author}{\bibinfo{person}{Kundan Krishna}, \bibinfo{person}{Sanjana Ramprasad}, \bibinfo{person}{Prakhar Gupta}, \bibinfo{person}{Byron~C. Wallace}, \bibinfo{person}{Zachary~C. Lipton}, {and} \bibinfo{person}{Jeffrey~P. Bigham}.} \bibinfo{year}{2025}\natexlab{}.
\newblock \bibinfo{title}{GenAudit: Fixing Factual Errors in Language Model Outputs with Evidence}.
\newblock
\showeprint[arxiv]{2402.12566}~[cs.CL]
\urldef\tempurl%
\url{https://arxiv.org/abs/2402.12566}
\showURL{%
\tempurl}


\bibitem[Laban et~al\mbox{.}(2024)]%
        {laban2024inksync}
\bibfield{author}{\bibinfo{person}{Philippe Laban}, \bibinfo{person}{Jesse Vig}, \bibinfo{person}{Marti Hearst}, \bibinfo{person}{Caiming Xiong}, {and} \bibinfo{person}{Chien-Sheng Wu}.} \bibinfo{year}{2024}\natexlab{}.
\newblock \showarticletitle{Beyond the Chat: Executable and Verifiable Text-Editing with LLMs}. In \bibinfo{booktitle}{\emph{Proceedings of the 37th Annual ACM Symposium on User Interface Software and Technology}} (Pittsburgh, PA, USA) \emph{(\bibinfo{series}{UIST '24})}. \bibinfo{publisher}{Association for Computing Machinery}, \bibinfo{address}{New York, NY, USA}, Article \bibinfo{articleno}{20}, \bibinfo{numpages}{23}~pages.
\newblock
\showISBNx{9798400706288}
\href{https://doi.org/10.1145/3654777.3676419}{doi:\nolinkurl{10.1145/3654777.3676419}}


\bibitem[Lee et~al\mbox{.}(2025)]%
        {lee2025aisurvey}
\bibfield{author}{\bibinfo{person}{Hao-Ping~(Hank) Lee}, \bibinfo{person}{Advait Sarkar}, \bibinfo{person}{Lev Tankelevitch}, \bibinfo{person}{Ian Drosos}, \bibinfo{person}{Sean Rintel}, \bibinfo{person}{Richard Banks}, {and} \bibinfo{person}{Nicholas Wilson}.} \bibinfo{year}{2025}\natexlab{}.
\newblock \showarticletitle{The Impact of Generative AI on Critical Thinking: Self-Reported Reductions in Cognitive Effort and Confidence Effects From a Survey of Knowledge Workers}. In \bibinfo{booktitle}{\emph{Proceedings of the 2025 CHI Conference on Human Factors in Computing Systems}} \emph{(\bibinfo{series}{CHI '25})}. \bibinfo{publisher}{Association for Computing Machinery}, \bibinfo{address}{New York, NY, USA}, Article \bibinfo{articleno}{1121}, \bibinfo{numpages}{22}~pages.
\newblock
\showISBNx{9798400713941}
\href{https://doi.org/10.1145/3706598.3713778}{doi:\nolinkurl{10.1145/3706598.3713778}}


\bibitem[Li et~al\mbox{.}(2023)]%
        {li2023halueval}
\bibfield{author}{\bibinfo{person}{Junyi Li}, \bibinfo{person}{Xiaoxue Cheng}, \bibinfo{person}{Xin Zhao}, \bibinfo{person}{Jian-Yun Nie}, {and} \bibinfo{person}{Ji-Rong Wen}.} \bibinfo{year}{2023}\natexlab{}.
\newblock \showarticletitle{{H}alu{E}val: A Large-Scale Hallucination Evaluation Benchmark for Large Language Models}. In \bibinfo{booktitle}{\emph{Proceedings of the 2023 Conference on Empirical Methods in Natural Language Processing}}, \bibfield{editor}{\bibinfo{person}{Houda Bouamor}, \bibinfo{person}{Juan Pino}, {and} \bibinfo{person}{Kalika Bali}} (Eds.). \bibinfo{publisher}{Association for Computational Linguistics}, \bibinfo{address}{Singapore}, \bibinfo{pages}{6449--6464}.
\newblock
\href{https://doi.org/10.18653/v1/2023.emnlp-main.397}{doi:\nolinkurl{10.18653/v1/2023.emnlp-main.397}}


\bibitem[Liu et~al\mbox{.}(2022)]%
        {liu2022hades}
\bibfield{author}{\bibinfo{person}{Tianyu Liu}, \bibinfo{person}{Yizhe Zhang}, \bibinfo{person}{Chris Brockett}, \bibinfo{person}{Yi Mao}, \bibinfo{person}{Zhifang Sui}, \bibinfo{person}{Weizhu Chen}, {and} \bibinfo{person}{Bill Dolan}.} \bibinfo{year}{2022}\natexlab{}.
\newblock \showarticletitle{A Token-level Reference-free Hallucination Detection Benchmark for Free-form Text Generation}. In \bibinfo{booktitle}{\emph{Proceedings of the 60th Annual Meeting of the Association for Computational Linguistics (Volume 1: Long Papers)}}, \bibfield{editor}{\bibinfo{person}{Smaranda Muresan}, \bibinfo{person}{Preslav Nakov}, {and} \bibinfo{person}{Aline Villavicencio}} (Eds.). \bibinfo{publisher}{Association for Computational Linguistics}, \bibinfo{address}{Dublin, Ireland}, \bibinfo{pages}{6723--6737}.
\newblock
\href{https://doi.org/10.18653/v1/2022.acl-long.464}{doi:\nolinkurl{10.18653/v1/2022.acl-long.464}}


\bibitem[Lo et~al\mbox{.}(2024)]%
        {lo2024semantic}
\bibfield{author}{\bibinfo{person}{Kyle Lo}, \bibinfo{person}{Joseph~Chee Chang}, \bibinfo{person}{Andrew Head}, \bibinfo{person}{Jonathan Bragg}, \bibinfo{person}{Amy~X. Zhang}, \bibinfo{person}{Cassidy Trier}, \bibinfo{person}{Chloe Anastasiades}, \bibinfo{person}{Tal August}, \bibinfo{person}{Russell Authur}, \bibinfo{person}{Danielle Bragg}, \bibinfo{person}{Erin Bransom}, \bibinfo{person}{Isabel Cachola}, \bibinfo{person}{Stefan Candra}, \bibinfo{person}{Yoganand Chandrasekhar}, \bibinfo{person}{Yen-Sung Chen}, \bibinfo{person}{Evie Yu-Yen Cheng}, \bibinfo{person}{Yvonne Chou}, \bibinfo{person}{Doug Downey}, \bibinfo{person}{Rob Evans}, \bibinfo{person}{Raymond Fok}, \bibinfo{person}{Fangzhou Hu}, \bibinfo{person}{Regan Huff}, \bibinfo{person}{Dongyeop Kang}, \bibinfo{person}{Tae~Soo Kim}, \bibinfo{person}{Rodney Kinney}, \bibinfo{person}{Aniket Kittur}, \bibinfo{person}{Hyeonsu~B. Kang}, \bibinfo{person}{Egor Klevak}, \bibinfo{person}{Bailey Kuehl}, \bibinfo{person}{Michael~J. Langan},
  \bibinfo{person}{Matt Latzke}, \bibinfo{person}{Jaron Lochner}, \bibinfo{person}{Kelsey MacMillan}, \bibinfo{person}{Eric Marsh}, \bibinfo{person}{Tyler Murray}, \bibinfo{person}{Aakanksha Naik}, \bibinfo{person}{Ngoc-Uyen Nguyen}, \bibinfo{person}{Srishti Palani}, \bibinfo{person}{Soya Park}, \bibinfo{person}{Caroline Paulic}, \bibinfo{person}{Napol Rachatasumrit}, \bibinfo{person}{Smita Rao}, \bibinfo{person}{Paul Sayre}, \bibinfo{person}{Zejiang Shen}, \bibinfo{person}{Pao Siangliulue}, \bibinfo{person}{Luca Soldaini}, \bibinfo{person}{Huy Tran}, \bibinfo{person}{Madeleine van Zuylen}, \bibinfo{person}{Lucy~Lu Wang}, \bibinfo{person}{Christopher Wilhelm}, \bibinfo{person}{Caroline Wu}, \bibinfo{person}{Jiangjiang Yang}, \bibinfo{person}{Angele Zamarron}, \bibinfo{person}{Marti~A. Hearst}, {and} \bibinfo{person}{Daniel~S. Weld}.} \bibinfo{year}{2024}\natexlab{}.
\newblock \showarticletitle{The Semantic Reader Project}.
\newblock \bibinfo{journal}{\emph{Commun. ACM}} \bibinfo{volume}{67}, \bibinfo{number}{10} (\bibinfo{date}{Sept.} \bibinfo{year}{2024}), \bibinfo{pages}{50–61}.
\newblock
\showISSN{0001-0782}
\href{https://doi.org/10.1145/3659096}{doi:\nolinkurl{10.1145/3659096}}


\bibitem[Mancini and Buckingham~Shum(2006)]%
        {ManciniBuckinghamShum2006}
\bibfield{author}{\bibinfo{person}{Clara Mancini} {and} \bibinfo{person}{Simon~J. Buckingham~Shum}.} \bibinfo{year}{2006}\natexlab{}.
\newblock \showarticletitle{Modelling discourse in contested domains: A semiotic and cognitive framework}.
\newblock \bibinfo{journal}{\emph{International Journal of Human-Computer Studies}} \bibinfo{volume}{64}, \bibinfo{number}{11} (\bibinfo{year}{2006}), \bibinfo{pages}{1154--1171}.
\newblock
\href{https://doi.org/10.1016/j.ijhcs.2006.07.002}{doi:\nolinkurl{10.1016/j.ijhcs.2006.07.002}}


\bibitem[Metropolitansky and Larson(2025a)]%
        {metropolitansky2025towards}
\bibfield{author}{\bibinfo{person}{Dasha Metropolitansky} {and} \bibinfo{person}{Jonathan Larson}.} \bibinfo{year}{2025}\natexlab{a}.
\newblock \showarticletitle{Towards Effective Extraction and Evaluation of Factual Claims}.
\newblock \bibinfo{journal}{\emph{arXiv preprint arXiv:2502.10855}} (\bibinfo{year}{2025}).
\newblock


\bibitem[Metropolitansky and Larson(2025b)]%
        {metropolitansky2025veritrail}
\bibfield{author}{\bibinfo{person}{Dasha Metropolitansky} {and} \bibinfo{person}{Jonathan Larson}.} \bibinfo{year}{2025}\natexlab{b}.
\newblock \showarticletitle{VeriTrail: Closed-Domain Hallucination Detection with Traceability}.
\newblock \bibinfo{journal}{\emph{arXiv preprint arXiv:2505.21786}} (\bibinfo{year}{2025}).
\newblock


\bibitem[Min et~al\mbox{.}(2023)]%
        {min2023factscore}
\bibfield{author}{\bibinfo{person}{Sewon Min}, \bibinfo{person}{Kalpesh Krishna}, \bibinfo{person}{Xinxi Lyu}, \bibinfo{person}{Mike Lewis}, \bibinfo{person}{Wen-tau Yih}, \bibinfo{person}{Pang Koh}, \bibinfo{person}{Mohit Iyyer}, \bibinfo{person}{Luke Zettlemoyer}, {and} \bibinfo{person}{Hannaneh Hajishirzi}.} \bibinfo{year}{2023}\natexlab{}.
\newblock \showarticletitle{Factscore: Fine-grained atomic evaluation of factual precision in long form text generation}. In \bibinfo{booktitle}{\emph{Proceedings of the 2023 Conference on Empirical Methods in Natural Language Processing}}. \bibinfo{pages}{12076--12100}.
\newblock


\bibitem[Murugadoss et~al\mbox{.}(2025)]%
        {murugadoss2025evaluating}
\bibfield{author}{\bibinfo{person}{Bhuvanashree Murugadoss}, \bibinfo{person}{Christian Poelitz}, \bibinfo{person}{Ian Drosos}, \bibinfo{person}{Vu Le}, \bibinfo{person}{Nick McKenna}, \bibinfo{person}{Carina~Suzana Negreanu}, \bibinfo{person}{Chris Parnin}, {and} \bibinfo{person}{Advait Sarkar}.} \bibinfo{year}{2025}\natexlab{}.
\newblock \showarticletitle{Evaluating the Evaluator: Measuring LLMs’ Adherence to Task Evaluation Instructions}.
\newblock \bibinfo{journal}{\emph{Proceedings of the AAAI Conference on Artificial Intelligence}} \bibinfo{volume}{39}, \bibinfo{number}{18} (\bibinfo{date}{Apr.} \bibinfo{year}{2025}), \bibinfo{pages}{19589--19597}.
\newblock
\href{https://doi.org/10.1609/aaai.v39i18.34157}{doi:\nolinkurl{10.1609/aaai.v39i18.34157}}


\bibitem[Nicholson et~al\mbox{.}(2021)]%
        {nicholson2021scite}
\bibfield{author}{\bibinfo{person}{Josh~M Nicholson}, \bibinfo{person}{Milo Mordaunt}, \bibinfo{person}{Patrice Lopez}, \bibinfo{person}{Ashish Uppala}, \bibinfo{person}{Domenic Rosati}, \bibinfo{person}{Neves~P Rodrigues}, \bibinfo{person}{Peter Grabitz}, {and} \bibinfo{person}{Sean~C Rife}.} \bibinfo{year}{2021}\natexlab{}.
\newblock \showarticletitle{scite: A smart citation index that displays the context of citations and classifies their intent using deep learning}.
\newblock \bibinfo{journal}{\emph{Quantitative science studies}} \bibinfo{volume}{2}, \bibinfo{number}{3} (\bibinfo{year}{2021}), \bibinfo{pages}{882--898}.
\newblock


\bibitem[Niu et~al\mbox{.}(2024)]%
        {niu2024ragtruth}
\bibfield{author}{\bibinfo{person}{Cheng Niu}, \bibinfo{person}{Yuanhao Wu}, \bibinfo{person}{Juno Zhu}, \bibinfo{person}{Siliang Xu}, \bibinfo{person}{Kashun Shum}, \bibinfo{person}{Randy Zhong}, \bibinfo{person}{Juntong Song}, {and} \bibinfo{person}{Tong Zhang}.} \bibinfo{year}{2024}\natexlab{}.
\newblock \showarticletitle{Ragtruth: A hallucination corpus for developing trustworthy retrieval-augmented language models}. In \bibinfo{booktitle}{\emph{Proceedings of the 62nd Annual Meeting of the Association for Computational Linguistics (Volume 1: Long Papers)}}. \bibinfo{pages}{10862--10878}.
\newblock


\bibitem[Pagnoni et~al\mbox{.}(2021)]%
        {pagnoni2021understanding}
\bibfield{author}{\bibinfo{person}{Artidoro Pagnoni}, \bibinfo{person}{Vidhisha Balachandran}, {and} \bibinfo{person}{Yulia Tsvetkov}.} \bibinfo{year}{2021}\natexlab{}.
\newblock \showarticletitle{Understanding Factuality in Abstractive Summarization with FRANK: A Benchmark for Factuality Metrics}. In \bibinfo{booktitle}{\emph{Proceedings of the 2021 Conference of the North American Chapter of the Association for Computational Linguistics: Human Language Technologies}}. \bibinfo{pages}{4812--4829}.
\newblock


\bibitem[Rachatasumrit et~al\mbox{.}(2022)]%
        {rachatasumrit2022citeread}
\bibfield{author}{\bibinfo{person}{Napol Rachatasumrit}, \bibinfo{person}{Jonathan Bragg}, \bibinfo{person}{Amy~X. Zhang}, {and} \bibinfo{person}{Daniel~S Weld}.} \bibinfo{year}{2022}\natexlab{}.
\newblock \showarticletitle{CiteRead: Integrating Localized Citation Contexts into Scientific Paper Reading}. In \bibinfo{booktitle}{\emph{Proceedings of the 27th International Conference on Intelligent User Interfaces}} (Helsinki, Finland) \emph{(\bibinfo{series}{IUI '22})}. \bibinfo{publisher}{Association for Computing Machinery}, \bibinfo{address}{New York, NY, USA}, \bibinfo{pages}{707–719}.
\newblock
\showISBNx{9781450391443}
\href{https://doi.org/10.1145/3490099.3511162}{doi:\nolinkurl{10.1145/3490099.3511162}}


\bibitem[Sarkar(2023)]%
        {sarkar2023aiknowledgework}
\bibfield{author}{\bibinfo{person}{Advait Sarkar}.} \bibinfo{year}{2023}\natexlab{}.
\newblock \showarticletitle{Exploring Perspectives on the Impact of Artificial Intelligence on the Creativity of Knowledge Work: Beyond Mechanised Plagiarism and Stochastic Parrots}. In \bibinfo{booktitle}{\emph{Proceedings of the 2nd Annual Meeting of the Symposium on Human-Computer Interaction for Work}} (Oldenburg, Germany) \emph{(\bibinfo{series}{CHIWORK '23})}. \bibinfo{publisher}{Association for Computing Machinery}, \bibinfo{address}{New York, NY, USA}, Article \bibinfo{articleno}{13}, \bibinfo{numpages}{17}~pages.
\newblock
\showISBNx{9798400708077}
\href{https://doi.org/10.1145/3596671.3597650}{doi:\nolinkurl{10.1145/3596671.3597650}}


\bibitem[Sarkar(2024a)]%
        {sarkar2024aiprovocateur}
\bibfield{author}{\bibinfo{person}{Advait Sarkar}.} \bibinfo{year}{2024}\natexlab{a}.
\newblock \showarticletitle{{AI Should Challenge, Not Obey}}.
\newblock \bibinfo{journal}{\emph{Commun. ACM}} (\bibinfo{date}{Sept.} \bibinfo{year}{2024}), \bibinfo{numpages}{5}~pages.
\newblock
\showISSN{0001-0782}
\href{https://doi.org/10.1145/3649404}{doi:\nolinkurl{10.1145/3649404}}
\newblock
\shownote{Online First}.


\bibitem[Sarkar(2024b)]%
        {sarkar2024llmscannotexplain}
\bibfield{author}{\bibinfo{person}{Advait Sarkar}.} \bibinfo{year}{2024}\natexlab{b}.
\newblock \showarticletitle{Large Language Models Cannot Explain Themselves}. In \bibinfo{booktitle}{\emph{Proceedings of the ACM CHI 2024 Workshop on Human-Centered Explainable AI}} (Honolulu, HI, USA) \emph{(\bibinfo{series}{HCXAI at CHI '24})}.
\newblock
\href{https://doi.org/10.48550/arXiv.2405.04382}{doi:\nolinkurl{10.48550/arXiv.2405.04382}}


\bibitem[Sarkar et~al\mbox{.}(2024)]%
        {sarkar2024genAIcritical}
\bibfield{author}{\bibinfo{person}{Advait Sarkar}, \bibinfo{person}{Xiaotong~(Tone) Xu}, \bibinfo{person}{Neil Toronto}, \bibinfo{person}{Ian Drosos}, {and} \bibinfo{person}{Christian Poelitz}.} \bibinfo{year}{2024}\natexlab{}.
\newblock \showarticletitle{{When Copilot Becomes Autopilot: Generative AI's Critical Risk to Knowledge Work and a Critical Solution}}. In \bibinfo{booktitle}{\emph{{Proceedings of the Annual Conference of the European Spreadsheet Risks Interest Group (EuSpRIG 2024)}}}.
\newblock


\bibitem[Sultanum and Srinivasan(2023)]%
        {sultanum2023datatales}
\bibfield{author}{\bibinfo{person}{Nicole Sultanum} {and} \bibinfo{person}{Arjun Srinivasan}.} \bibinfo{year}{2023}\natexlab{}.
\newblock \showarticletitle{DATATALES: Investigating the use of Large Language Models for Authoring Data-Driven Articles}. In \bibinfo{booktitle}{\emph{2023 IEEE Visualization and Visual Analytics (VIS)}}. \bibinfo{pages}{231--235}.
\newblock
\href{https://doi.org/10.1109/VIS54172.2023.00055}{doi:\nolinkurl{10.1109/VIS54172.2023.00055}}


\bibitem[Tankelevitch et~al\mbox{.}(2024)]%
        {tankelevitch2024GenAImetacognition}
\bibfield{author}{\bibinfo{person}{Lev Tankelevitch}, \bibinfo{person}{Viktor Kewenig}, \bibinfo{person}{Auste Simkute}, \bibinfo{person}{Ava~Elizabeth Scott}, \bibinfo{person}{Advait Sarkar}, \bibinfo{person}{Abigail Sellen}, {and} \bibinfo{person}{Sean Rintel}.} \bibinfo{year}{2024}\natexlab{}.
\newblock \showarticletitle{The Metacognitive Demands and Opportunities of Generative AI}. In \bibinfo{booktitle}{\emph{Proceedings of the CHI Conference on Human Factors in Computing Systems}} (Honolulu, HI, USA) \emph{(\bibinfo{series}{CHI '24})}. \bibinfo{publisher}{Association for Computing Machinery}, \bibinfo{address}{New York, NY, USA}, Article \bibinfo{articleno}{680}, \bibinfo{numpages}{24}~pages.
\newblock
\showISBNx{9798400703300}
\href{https://doi.org/10.1145/3613904.3642902}{doi:\nolinkurl{10.1145/3613904.3642902}}


\bibitem[Toulmin(2003)]%
        {Toulmin2003-xj}
\bibfield{author}{\bibinfo{person}{Stephen~E Toulmin}.} \bibinfo{year}{2003}\natexlab{}.
\newblock \bibinfo{booktitle}{\emph{The Uses of Argument}}.
\newblock \bibinfo{publisher}{Cambridge University Press}, \bibinfo{address}{Cambridge, England}.
\newblock


\bibitem[Uren et~al\mbox{.}(2006)]%
        {UrenBuckinghamShum2006}
\bibfield{author}{\bibinfo{person}{Victoria Uren}, \bibinfo{person}{Simon Buckingham~Shum}, \bibinfo{person}{Michelle Bachler}, {and} \bibinfo{person}{Gangmin Li}.} \bibinfo{year}{2006}\natexlab{}.
\newblock \showarticletitle{Sensemaking tools for understanding research literatures: design, implementation and user evaluation}.
\newblock \bibinfo{journal}{\emph{International Journal of Human-Computer Studies}} \bibinfo{volume}{64}, \bibinfo{number}{5} (\bibinfo{year}{2006}), \bibinfo{pages}{420--445}.
\newblock
\href{https://doi.org/10.1016/j.ijhcs.2005.09.004}{doi:\nolinkurl{10.1016/j.ijhcs.2005.09.004}}


\bibitem[Yan et~al\mbox{.}(2025)]%
        {yan2025answering}
\bibfield{author}{\bibinfo{person}{Litao Yan}, \bibinfo{person}{Jeffrey Tao}, \bibinfo{person}{Lydia~B Chilton}, {and} \bibinfo{person}{Andrew Head}.} \bibinfo{year}{2025}\natexlab{}.
\newblock \showarticletitle{Answering Developer Questions with Annotated Agent-Discovered Program Traces}. In \bibinfo{booktitle}{\emph{Proceedings of the 38th Annual ACM Symposium on User Interface Software and Technology}} \emph{(\bibinfo{series}{UIST '25})}. \bibinfo{publisher}{Association for Computing Machinery}, \bibinfo{address}{New York, NY, USA}, Article \bibinfo{articleno}{29}, \bibinfo{numpages}{14}~pages.
\newblock
\showISBNx{9798400720376}
\href{https://doi.org/10.1145/3746059.3747652}{doi:\nolinkurl{10.1145/3746059.3747652}}


\bibitem[Yue et~al\mbox{.}(2023)]%
        {yue2023attrscore}
\bibfield{author}{\bibinfo{person}{Xiang Yue}, \bibinfo{person}{Boshi Wang}, \bibinfo{person}{Ziru Chen}, \bibinfo{person}{Kai Zhang}, \bibinfo{person}{Yu Su}, {and} \bibinfo{person}{Huan Sun}.} \bibinfo{year}{2023}\natexlab{}.
\newblock \showarticletitle{Automatic Evaluation of Attribution by Large Language Models}. In \bibinfo{booktitle}{\emph{Findings of the Association for Computational Linguistics: EMNLP 2023}}, \bibfield{editor}{\bibinfo{person}{Houda Bouamor}, \bibinfo{person}{Juan Pino}, {and} \bibinfo{person}{Kalika Bali}} (Eds.). \bibinfo{publisher}{Association for Computational Linguistics}, \bibinfo{address}{Singapore}, \bibinfo{pages}{4615--4635}.
\newblock
\href{https://doi.org/10.18653/v1/2023.findings-emnlp.307}{doi:\nolinkurl{10.18653/v1/2023.findings-emnlp.307}}


\bibitem[Zhang et~al\mbox{.}(2025)]%
        {zhang2025treereader}
\bibfield{author}{\bibinfo{person}{Zijian Zhang}, \bibinfo{person}{Pan Chen}, \bibinfo{person}{Fangshi Du}, \bibinfo{person}{Runlong Ye}, \bibinfo{person}{Oliver Huang}, \bibinfo{person}{Michael Liut}, {and} \bibinfo{person}{Alán Aspuru-Guzik}.} \bibinfo{year}{2025}\natexlab{}.
\newblock \showarticletitle{TreeReader: A Hierarchical Academic Paper Reader Powered by Language Models}. In \bibinfo{booktitle}{\emph{2025 IEEE Symposium on Visual Languages and Human-Centric Computing (VL/HCC)}}. \bibinfo{pages}{286--292}.
\newblock
\href{https://doi.org/10.1109/VL-HCC65237.2025.00039}{doi:\nolinkurl{10.1109/VL-HCC65237.2025.00039}}


\bibitem[Zheng et~al\mbox{.}(2023)]%
        {zheng2023judging}
\bibfield{author}{\bibinfo{person}{Lianmin Zheng}, \bibinfo{person}{Wei-Lin Chiang}, \bibinfo{person}{Ying Sheng}, \bibinfo{person}{Siyuan Zhuang}, \bibinfo{person}{Zhanghao Wu}, \bibinfo{person}{Yonghao Zhuang}, \bibinfo{person}{Zi Lin}, \bibinfo{person}{Zhuohan Li}, \bibinfo{person}{Dacheng Li}, \bibinfo{person}{Eric Xing}, {et~al\mbox{.}}} \bibinfo{year}{2023}\natexlab{}.
\newblock \showarticletitle{Judging llm-as-a-judge with mt-bench and chatbot arena}.
\newblock \bibinfo{journal}{\emph{Advances in neural information processing systems}}  \bibinfo{volume}{36} (\bibinfo{year}{2023}), \bibinfo{pages}{46595--46623}.
\newblock


\end{thebibliography}

\end{document}